\documentstyle[epsf,html]{l-aa}
\def\gapprox{\;\rlap{\lower 2.5pt            
 \hbox{$\sim$}}\raise 1.5pt\hbox{$>$}\;}       
\def\lapprox{\;\rlap{\lower 2.5pt            
 \hbox{$\sim$}}\raise 1.5pt\hbox{$<$}\;} 
\def\N{\,{\rm I\kern-.20em N}}
\begin{document}
\thesaurus{02.01.1; 06.03.2; 06.06.3; 06.13.1; 06.18.1}
\title{High-sensitivity observations of solar flare decimeter radiation}
\author{Arnold O. Benz
\and   Peter Messmer
\and Christian Monstein}
\offprints{ A.O. Benz, benz@astro.phys.ethz.ch}
\institute{Institute of Astronomy, ETH-Zentrum, CH-8092 Zurich, Switzerland}
\date{Received   ........ ; accepted ........ }
\maketitle
\markboth{Solar flare decimeter radiation}{}
\begin{abstract}
A new acousto-optic radio spectrometer has observed the 1 -- 2 GHz radio emission 
of solar flares with unprecedented sensitivity. The number of detected decimeter 
type III bursts is greatly enhanced compared to observations by conventional 
spectrometers observing only one frequency at the time. The observations 
indicate a large number of electron beams propagating in dense plasmas. For the first time, we report weak, reversed drifting type III bursts at {\sl frequencies above} simultaneous narrowband 
decimeter spikes. The type III bursts are reliable signatures of electron beams 
propagating downward in the corona, apparently away from the source of the 
spikes. The observations contradict the most popular spike model that places the spike sources at the footpoints of loops. Conspicuous also was an apparent bidirectional type U burst forming a fish-like pattern. It occurs simultaneously with an intense U-burst at 600-370 MHz observed in Tremsdorf. We suggest that it intermodulated with strong {\sl terrestrial interference} (cellular phones) causing a spurious symmetric pattern in the spectrogram at 1.4 GHz. Symmetric features in the 1 -- 2 GHz range, some already reported in the literature, therefore must be considered with utmost caution.
\end{abstract} 
\hspace{1cm}\\

\section{Introduction}
The acceleration of a large population of superthermal electrons is a 
characteristic of solar flares. The acceleration process is still unclear, 
although the original energy release is generally, but not unanimously believed 
to be due to reconnection (e.g. reviews by Tandberg-Hanssen \& Emslie 1988; 
Priest \& Forbes 2000). Most of these particles escape from the acceleration region along 
magnetic field lines. If the distance of propagation is long enough, a 
bump-on-tail instability develops driving Langmuir waves. Thus one may expect 
plenty of radio emission of type III in every flare. This is not the case as 
pointed out e.g. by Simnett \& Benz (1986). There may be several reasons 
conspiring: {\sl (i)} Observations were made mostly below 1 GHz 
corresponding to densities below a few times $10^9$cm$^{-3}$, as was the case in 
the above mentioned study and in earlier work. However, flare particle 
acceleration and propagation is suspected to generally occur at higher densities 
(Miller et at. 1997). {\sl (ii)} Even harmonic radio emission (at twice the 
plasma frequency) is increasingly free-free absorbed at higher frequencies, thus 
the observed type III bursts appear weaker. {\sl (iii)} The thermal background  
emission of the Sun increases strongly with frequency around 1 GHz, making burst
detection even more difficult. 

Magnetic fields at low coronal altitude and particularly in active regions are 
well known to have predominantly loop-like shapes. Thus electron beams cannot 
easily escape into interplanetary space. Electron beams radiating type III 
bursts at decimeter frequencies must follow such loops and thus have a short total lifetime. Due to the smallness of the loops and small density range, they have often a small bandwidth (Benz et al. 1983). Also, the instantaneous duration of bursts decreases with frequency about in inverse proportion (Staehli \& Benz 1987; Mel\'endez et al. 1999). Electron beams moving into the direction of higher magnetic field and density emit less radio emission (Robinson \& Benz 2000). Small bandwidth, short duration, and downward motion enhance the difficulty for observations even further. 

On the other hand, the observation of high-frequency type III bursts is very attractive as electron beams following magnetic field lines map the magnetic geometry into the frequency-time plane (i.e. the spectrogram). Decimeter type III bursts thus outline magnetic fields related to reconnection. 

Meterwave Type III bursts have been reported to be associated by narrowband spikes above the starting frequencies in 10\% of all cases (Benz et al. 1983). These type III emissions are regular drifting bursts, representing electron beams propagating toward lower density, i.e. upward in the corona. Groups of {\sl meterwave} spikes are well correlated with individual type III bursts (Benz et al. 1996). Metric spikes usually occur in a relatively small band of frequencies around 300 MHz. Upward drifting type III bursts have also been noted in practically all events of {\sl decimeter} spikes (Benz \& G\"udel 1987). Such associated type III bursts occur in the meter waves, presumably at higher altitude and are often poorly correlated with time structures of the spike group. Narrowband spikes at meter and decimeter wavelength have similar properties concerning individual bursts, including duration, bandwidth and polarization (review by Benz 1986). However, decimeter spikes form much richer groups extending over a wider frequency band and are well associated with hard X-ray events (Benz \& Kane 1986; G\"udel et al. 1989). 

Reversed, downward drifting type III bursts have never been noted in association with any spikes. This seems to suggest that the sources of spike emission are located close to the bottom of the corona, near the region where energetic electrons are lost from the trap. It has generally been taken as evidence for spikes to occur at the footpoints of loops, where the velocity distribution of trapped electrons has a loss-cone distribution and is unstable toward the generally favored growing upper hybrid, Bernstein or electron cyclotron waves (Melrose \& Dulk 1982; Sharma et al. 1982; cf. recent theoretical work by Conway \& Willes  2000). In the alternative scenario, on the other hand, suggesting that spikes are a direct signature of the acceleration process and their sources therefore at some finite altitude, it is not clear why reversed drifting type III bursts indicating downward drifting electron beams have never been observed.

Here we report on observations with a multichannel spectrometer far exceeding the 
sensitivity of the usual frequency-agile instruments. We concentrate on the analysis of two particularly interesting type III events.

\section{Instrument and Observations}
Observations with an {\sl acousto-optic} spectrometer (AOS) of the type Elson-E2 
were made in Bleien (Switzerland) from February 15 to March 15, 2000 in the 1.0 
-- 2.0 GHz range.  A total number of 256 channels with a bandwidth of 1 MHz and a 
channel separation of 3.9 MHz were simultaneously observed. The integration 
and sampling times were equal and amounted to 40 ms. The spectrometer was fed by a 7 meter 
dish observing the full Sun. 

A second spectrometer, the {\sl frequency-agile} Phoenix-2 instrument (Messmer 
et al. 1999), observed in parallel with the same antenna. Its frequency range 
was from 1.0 to 2.8 GHz in 40 channels having a bandwidth of 10 MHz each and a 
channel separation of 40 MHz. The integration time was 0.44 ms, and the sampling 
time 20 ms. The observations of Phoenix-2 were used to calibrate the AOS data in 
absolute time and in flux density.

The improvement by the AOS for individual measurements is a factor of 3.1, in 
agreement with the radiometer equation. Compared to the sample size of 
Phoenix-2 in frequency and time (40 MHz $\times$ 20 ms), the reduction in rms 
noise is a factor of 7.1. 

The two spectrometers operated in parallel for about 250 hours. We searched for 
radio emission at the time of flares as given by the solar event list of NOAA (2000)
and noted radio emission in the Phoenix-2 data during a total number of 14 solar 
flares. The bursts included 6 patches of continuum emission, 1 pulsation, two 
synchrotron emissions and 5 type III bursts. The registrations by the AOS added 
the following information:

\begin{itemize}
\item Every radio event of any kind showed type III emission in AOS 
data in addition. The only exception to this rule was one broadband radiation probably 
emitted by the gyro-synchrotron mechanism.
\item Flares reported by NOAA without radio emission in the 1 -- 2 GHz band and not 
detected in Phoenix-2 data often were accompanied by short, narrowband type III 
emission in AOS data. This was not the case though for every small flare ($\leq$ 
C-class).
\end{itemize}

The first analysis of the data suggests as the main result of higher sensitivity 
the registration of a larger number of small type III bursts. The total number 
of such bursts in the AOS data is limited both by sensitivity as well as by possible confusion 
with interference. In some cases of short and narrowband bursts, it was not 
possible to recognize a drift in frequency and time, the characteristic for type III 
bursts. Without this signature, the classification of the burst type and even 
the identification as a solar event is not possible beyond doubt. 

The frequency distribution in peak flux of decimetric type III bursts is not known. At kilometer wavelength Fitzenreiter et al. (1976) found a power-law distribution with an exponent between -1.2 and -1.7. Aschwanden et al.(1998) find exponents in the same range at meter wavelengths. Assuming a power-law distribution with an exponent of -1.4, a sensitivity increase by a factor 7 should result in a larger burst number in AOS data than Phoenix-2 by a factor of 2.2.  The observed increase is a factor of 4.5. Several reasons may be responsible for this, including better burst recognition by enhanced frequency resolution, bias in the search procedure or a change in the distribution at low fluxes.

\begin{figure}
\begin{center}
\leavevmode
\vskip1.6cm
\mbox{\hspace{-3.2cm}\vspace{0.9cm}\epsfxsize=6.2cm\epsfbox{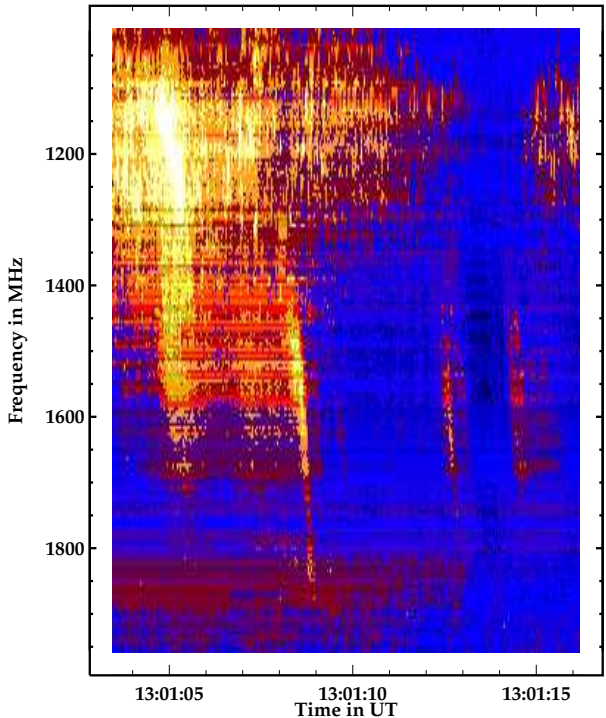}}
\end{center}
\vskip-1.0cm
\caption[]{Radio spectrogram recorded on 2000 February 22 by the AOS. At the 
top of the figure (low frequencies), narrowband spikes are visible up to about 
1400 MHz. Between 1400 and 1800 MHz at least 5 reversed drifting type III bursts 
are discernible.}
\end{figure} 
\section{Spikes and Type III Bursts}

Decimeter type III bursts observed at peak flux densities between 8 and 18 
sfu are presented in Fig. 1. For the first time, we report such bursts occurring 
at {\sl higher} frequencies than decimeter narrowband spikes. The spikes around 1200 MHz were also recorded by the frequency-agile spectrometer, but the type III bursts become only marginally noticeable after appropriate integration. The temporal correlation between spike activity and type III bursts is not in detail. The two emissions are well separated in frequency at about 1400 MHz. The drift rates of the decimeter type III bursts are +570$\pm 40$ MHz/s, thus reversed and in the range typical at the frequency (Mel\'endez et al. 1999). The positive sign indicates that the 
radiating electron beams propagate to higher density, i.e. downward in the corona. Assuming that type III bursts and spikes are emitted at similar frequencies (as suggested by the close correlation of meterwave spikes and type III bursts, Benz et al. 1996), the observed type III emission most likely originates at lower altitude than the spike source. This makes it difficult to interpret the spike emission as originating from the footpoints of loops, contrary to the most favored scenario (cf. section 1).

The Tremsdorf Observatory has recorded a faint regular-drift type III burst at 13:00:48 UT on 200 MHz. Even fainter type III-like activity occurred at 13:01:06 UT around 600 MHz. The San Vito radio flux meters report a peak value of only 60 sfu at 245 MHz (NOAA 2000). Thus the signatures of upward propagating electron beams are about one order of magnitude stronger than the emission of the downward beams. This is consistent with the previous non-detection of the latter.

We also note that the regular type III bursts (in the Tremsdorf data) and the reversed drift type IIIs (in the AOS data) are not symmetrical. This we interpret as evidence that the reversed-drifting structures are very likely not of instrumental origin (see below). 

\section{New Type of Interference}
\begin{figure}
\begin{center}
\leavevmode
\vskip1.0cm
\mbox{\hspace{-1.9cm}\vspace{5cm}\epsfxsize=6.8cm\epsfbox{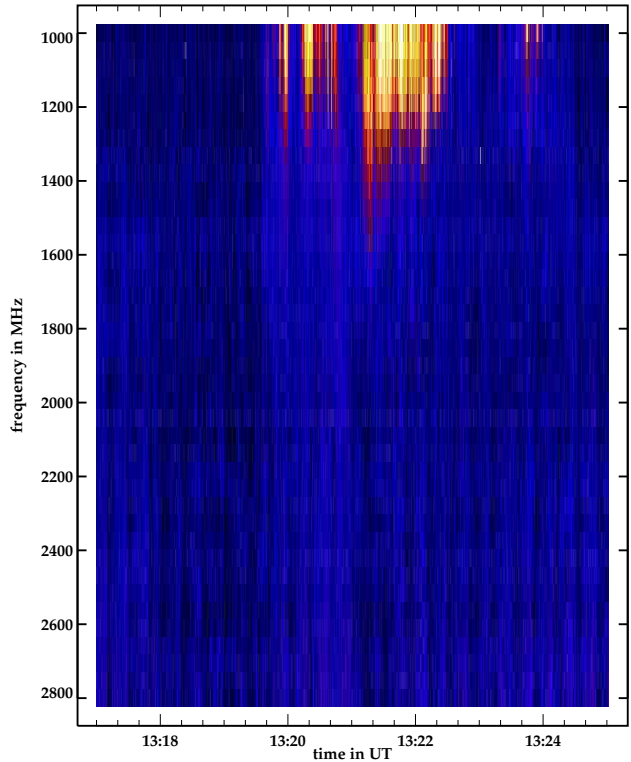}}
\vskip1.5cm
\mbox{\hspace{-2.2cm}\vspace{3cm}\epsfxsize=6.8cm\epsfbox{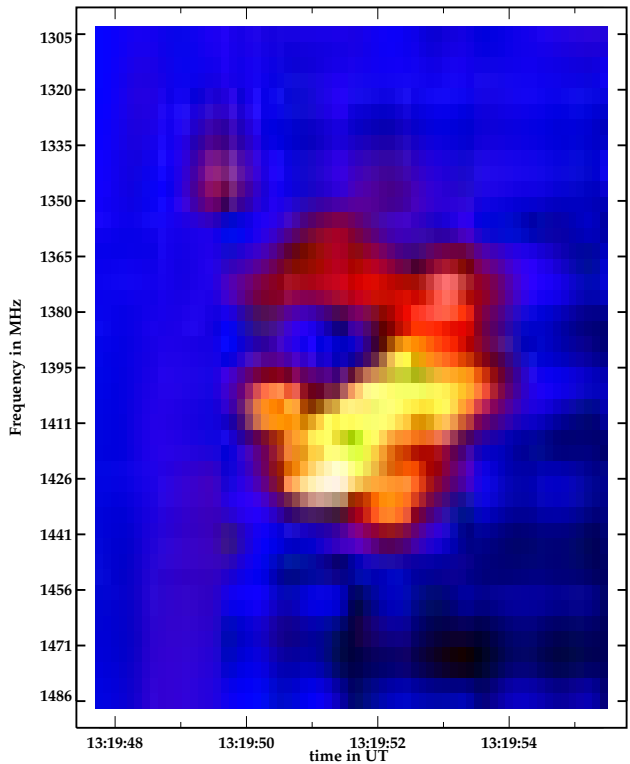}}
\end{center}
\vskip-1.0cm
\caption[]{{\bf a} Radio spectrogram recorded by the frequency-agile Phoenix-2, the standard 
instrument, giving an overview of the solar decimeter radiation in the 1 -- 2.8 GHz 
range.  {\bf b} An enlargement of apparent type III activity as observed by 
the temporary multichannel receiver (AOS).}
\end{figure} 

Fig. 2 displays an unusual type III-like event. The overview observed with the frequency-agile spectrometer is shown in Fig. 2a. The burst apparent in Fig. 2b, recorded by the multi-channel instrument, is not visible in the overview. Simultaneously and later, decimeter pulsations have been observed by both spectrometers. Their low-sensitivity recording is shown in Fig. 2a. Pulsating continuum emission has recently been modeled by Kliem et al. (2000), suggesting repeated secondary reconnection of magnetic islands in a current sheet. 

Fig. 2b shows slow-drifting emission, not visible in low-resolution (Fig 
2a). There is a distinct symmetry in frequency with an axis at 1392 MHz, 
forming a fish-shaped structure. The peak flux at the resolution of AOS was 62.5 
sfu, but most of the upper branch is around 10 sfu. The rms noise is 1.2 sfu. 
The duration of individual bursts at intermediate frequencies is $\leq$ 1 
second. 

The upper part of the event resembles a burst of subtype U. Type III bursts with 
U-shape (inverted in the representation of the frequency axis in Fig. 2b) are 
well known in the meter wavelength range and are generally interpreted as emissions of 
electron beams following magnetic loops in the corona (e.g. review by Benz 1993; 
Aurass \& Klein 1997). In the 1 -- 2 GHz range, U-bursts were observed with a total 
duration between 4 -- 5 seconds by Aschwanden et al. (1992). The inverse of a type
U burst has never been reported, although a somewhat similar structure, called 
Y-bursts, has been noticed by Tarnstrom \& Zehntner (1975) and interpreted by 
magnetic reflection of the electron beam in a converging magnetic field. Also, 
rising branches after U-bursts have been reported by Caroubalos et al.(1987) 
who introduced the term "N-burst". Simultaneous U-bursts and Y-bursts have never 
been observed.

\begin{figure}
\begin{center}
\leavevmode
\mbox{\hspace{0.2cm}\vspace{5cm}\epsfxsize=8.5cm\epsfbox{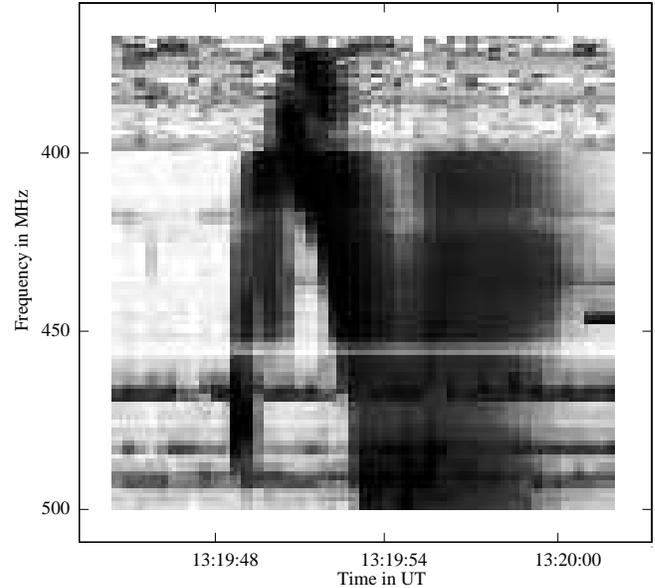}}
\end{center}
\vskip-0.5cm
\caption[]{Spectrogram observed by the Tremsdorf observatory (courtesy A. Klassen and G. Mann). Enhanced emission is shown dark.}
\end{figure} 

The fish-like structure in the spectrogram (Fig. 2b) is extremely symmetric 
and the up and down branches start within 0.1 s. It may be suggestive of a 
simultaneous injection of bidirectional beams. Several geometries are possible including propagation into two similar magnetic islands (plasmoids). Thus a scenario is possible where acceleration by secondary tearing occurs in the center of an 8-shaped magnetic field as predicted by island formation in reconnecting current sheets. 

In the following, we suggest that the fish-like structure in Fig. 2b is more likely caused by intermodulation of a solar radio burst with terrestrial interference. At the time of the emission feature, San Vito (NOAA 2000) reports an extremely strong radio burst of 10000 sfu at 410 MHz. Considering the arbitrariness of the sampling frequency and the low time resolution (1 s) of the flux meter, the reported peak value must be a lower limit. At 606 MHz RBR reports a peak flux of 45 sfu (NOAA 2000).

Figure 3 displays the spectrogram of Tremsdorf observatory in the range 370 -- 500 MHz. The data is not calibrated, and the background is not subtracted. At 400 MHz the sensitivity of the instrument changes. A very intense U-burst dominates the picture. 

Throughout the observations, strong terrestrial interference was observed by the AOS at 1827($\pm$4) MHz (not shown in Fig. 2b, omitted by Phoenix-2). Even stronger interference is usually present at 995($\pm$4) MHz, well known from previous observations, but not observed in the present case. Both interferences are caused by transmitters for cellular phones. Intermodulation is possible by a non-linear element in the chain of propagation and reception (most likely an amplifier). It produces spurious signals at
\begin{equation}
\omega \ =\ \omega_i \pm \omega_s\ \ \ ,
\end{equation}
where $\omega$ is the observing frequency, $\omega_i$ the frequency of the interference, and $\omega_s$ the solar signal.

Eq.(1) predicts two features in a spectrogram symmetric to the frequency of interference. Such structures cannot be perceived in the data of the frequency-agile spectrometer (Fig.2a). The band of the AOS is such that it does not include both sides of the symmetric spurious patterns for neither of the two interferences.

Symmetric features can also appear in the presence of two interferences $\omega_1$ and $\omega_2$ if intermodulation occurs for $\omega_1 + \omega_s$ and $\omega_2 - \omega_s$. A symmetric structure $\omega_0 \pm \omega_s^\star$ may thus form, where the frequency of symmetry, $\omega_0$, is given by
\begin{equation}
\omega_0 \ =\  {{\omega_1 + \omega_2}\over 2}\ \approx\ 1391(\pm 4)\ \ {\rm MHz}\ \
\end{equation}
for the given interference frequencies and
\begin{equation}
\omega_s^\star \ =\ \omega_s + \omega_1 - \omega_0\ =\ \omega_s - \omega_2 + \omega_0 \ \ \ .
\end{equation}
There is no steady emission at $\omega_0$. The spurious structure produced by the stronger interference appears more intense. These properties including the approximate value in Eq.(2) all agree well with the observed features in Fig. 2b.

The agreement between the original image in the Tremsdorf spectrogram (Fig.3) and its possible projections into the 1-2 GHz band of the AOS by interferences at 955 and 1827 MHz is reasonably good. The original U-burst has a range from 370 MHz to 420 MHz at the time of turnover. The predicted values for intermodulation with the 1827 MHz interference, 1407 and 1457 MHz respectively, include the observed range from 1407 to 1442 MHz. Intermodulation at 955 MHz would cause features between 1325 and 1375 MHz, while the observed extremes are 1353 and 1373 MHz, respectively.

The timing is more difficult to compare, as the U-burst in the Tremsdorf spectrogram is followed by a weak continuum (possibly to be classified as a decimetric type V burst), and the timing accuracy and resolution of the Tremsdorf image are about 0.3s. Nevertheless, the timing agrees within the uncertainties, and the total duration of the U-burst of 5.5s well contains the fish-like structure in the AOS data lasting 4.5s.

Therefore, the very likely interpretation of the apparent bidirectional U-burst in the AOS data (Fig.2b) is a spurious signal produced by intermodulation. The most intense part of a strong U-burst is mirrored at two frequencies of strong interference and shifted up in frequency to 1.4 GHz. The flux density of the intermodulation pattern is at least 3 orders of magnitude weaker than the original solar signal.

The identification with intermodulation reveals a serious threat of high-sensitivity broadband spectroscopy by strong terrestrial interference. For a reliable detection of a weak signal at decimeter waves, the possibility of a mirror image of a meterwave burst must be excluded. This puts some doubts for example on bidirectional type III bursts recently reported in the 1 -- 2 GHz range (Ning et al. 2000, Fig. 1) spreading out symmetrically from a frequency of 1720 MHz obviously interfered.

\section{Conclusions}
Enhanced sensitivity in the decimeter range greatly enlarges the number of 
detected type III bursts. As these bursts have a relatively small bandwidth and 
duration, high resolution in frequency and time is equally important. Spurious signals from intermodulation may be avoided technically i.e. by better amplifiers or excluded by monitoring the meter wavelength range of solar bursts. 

The bursts trace out paths of unstable electron beams propagating away from the 
acceleration region. One month of observations has immediately produced exciting 
results, including the first case of reversed-drift type III bursts starting at the frequencies of a group of narrowband decimeter spikes. This observation puts doubts on the standard model of decimetric spikes occurring in the lowest parts of loops, but needs confirmation by similar cases. 

More high-sensitivity observations are necessary for a complete overview on the 
activity of weak decimeter type III bursts. A systematic high-sensitivity survey in the frequency range 1 -- 3 GHz, possibly combined with spatial resolution, have the potential to generate a 
clearer understanding of magnetic fields associated with the acceleration region 
of electrons in flares.

\begin{acknowledgements} 
We thank Michael Arnold (ETH Zurich) and Max W\"uthrich (University of Bern) for help in the realization of the temporary  set-up. The acousto-optic spectrometer was kindly lent by the Microwave Department of the Institute of Applied Physics, University of Bern. Observations of the Tremsdorf spectrometer were essential in identifying the cause of the intermodulation pattern. They were kindly provided by A. Klassen and G. Mann. The work at ETH Zurich is financially supported by the Swiss National Science Foundation (grant No. 2000-061559.00).

\end{acknowledgements}

\end{document}